\pdfoutput=1
\documentclass{PoS}
\usepackage{sssmacros}
\usepackage{subfigure}
\def\mathunit#1{\mathrel{\mathrm{#1}}\mathclose{}\mathord{}}

\def\figref#1{Fig.~\ref{#1}}
\def\Figref#1{Figure~\ref{#1}}
\def\tabref#1{Table~\ref{#1}}
\def\fTR{\ensuremath{f_{\textrm{\scriptsize TR}}}}
\def\nBL{\ensuremath{n_{\textrm{\scriptsize BL}}}}
\def\statsyst#1#2{\pm#1\ \textnormal{(stat.)}\pm#2\ \textnormal{(syst.)}}

\long\def\threeboxesgap#1#2#3#4{%
  \setlength\figsize{\hsize}%
  \addtolength\figsize{-#4}%
  \addtolength\figsize{-#4}%
  \divide\figsize by 3
  \vbox{%
  \makebox{\parbox[t]{\figsize}{\vskip 0.1pt #1}%
           \hspace{#4}%
           \parbox[t]{\figsize}{\vskip 0.1pt #2}%
           \hspace{#4}%
           \parbox[t]{\figsize}{\vskip 0.1pt #3}}}}
\long\def\threeboxes#1#2#3{\threeboxesgap{#1}{#2}{#3}{\columnsep}}

\makeatletter
\def\switchtotable{\def\@captype{table}}
\def\switchtofigure{\def\@captype{figure}}
\makeatother

\title{Electron and Photon Performance and Electron $\pt$ Spectrum Measurement with ATLAS in $pp$ Collisions at $\sqrt{s} =7\tev$}

\ShortTitle{Electron and Photon Performance with ATLAS}

\author{\speaker{Scott Snyder}\\
        (For the ATLAS Collaboration)\\
	Brookhaven National Laboratory (BNL)\\
	E-mail: \email{snyder@bnl.gov}}

\abstract{We discuss the early performance of the reconstruction
  of electrons and photons
  at ATLAS with a center-of-mass energy of
  $7\tev$.  Using a data sample of about $15\inb$, we present a
  measurement of the transverse momentum distribution of
  inclusive electrons, as well as an observation of prompt
  photons and a measurement of the purity of this sample. We
  also demonstrate the reconstruction of $J/\psi$ mesons with a data
  sample of $78\inb$.}

\FullConference{35th International Conference of High Energy Physics\\
                 July 22-28, 2010\\
                 Paris, France}

\begin{document}

\section{Introduction}

Efficient reconstruction of electromagnetic (EM) objects, electrons and
photons, with a small background is crucial to many
of the physics results of interest at the LHC.  This note reports 
some early results on EM object reconstruction with
ATLAS at $\sqrt{s}=7\tev$, including
a measurement of the inclusive electron
$\pt$ spectrum and the observation of prompt photons and
$J/\psi$ mesons.


ATLAS is fully described in \cite{atlasdetector}; here, we draw attention
to a few relevant details.
The innermost pixel layer of the central tracker is at $r\sim5\cm$
and is useful for rejecting conversions.  The outermost tracking
detector is the transition
radiation tracker (TRT), consisting of straw tubes interleaved with
polypropylene radiators; it provides good discrimination
between electrons and pions of energies $1$--$200\gev$.
The EM calorimeter uses Pb/liquid Ar
with an accordion geometry and fine segmentation.
Within $|\eta|<2.5$, there are
three longitudinal layers; the first (layer 1) is
finely segmented in $\eta$ (up to 0.003), allowing precise
$\eta$ measurements and rejection of $\pi^0$ decays.

EM object reconstruction starts by finding seed clusters in the EM calorimeter
with significant energy; these can be defined either with a
($\eta$,$\phi$) window or with a nearest-neighbor clustering algorithm.
They are matched with tracks; on the basis of this,
each candidate is classified as either an electron, a photon, or a
converted photon.  A final cluster is formed from cells in a rectangular
region around the seed,
and the measurements from the tracker and calorimeter are combined.

\section{Inclusive electron analysis}

This analysis \cite{inclele} measures
the inclusive $\pt$ spectrum of electron candidates and
breaks down the contributions to this by source,
either $b/c\ra e$ ($Q$), conversions ($\gamma$), or hadrons ($h$).
We require $\et>7\gev$, $|\eta|<2.0$ and
make additional requirements on $f_1$, the fractional energy
in layer~1; the shower width and shape in layer~1;
the track's impact parameter with respect
to the event vertex and number of hits; and
$\Delta\eta(\textrm{track},\textrm{cluster})$.
This selection yields 67124 events in $13.8\inb$.

\begin{figure}
  \threeboxes
  {\includegraphics[width=\figsize,height=3.2cm]{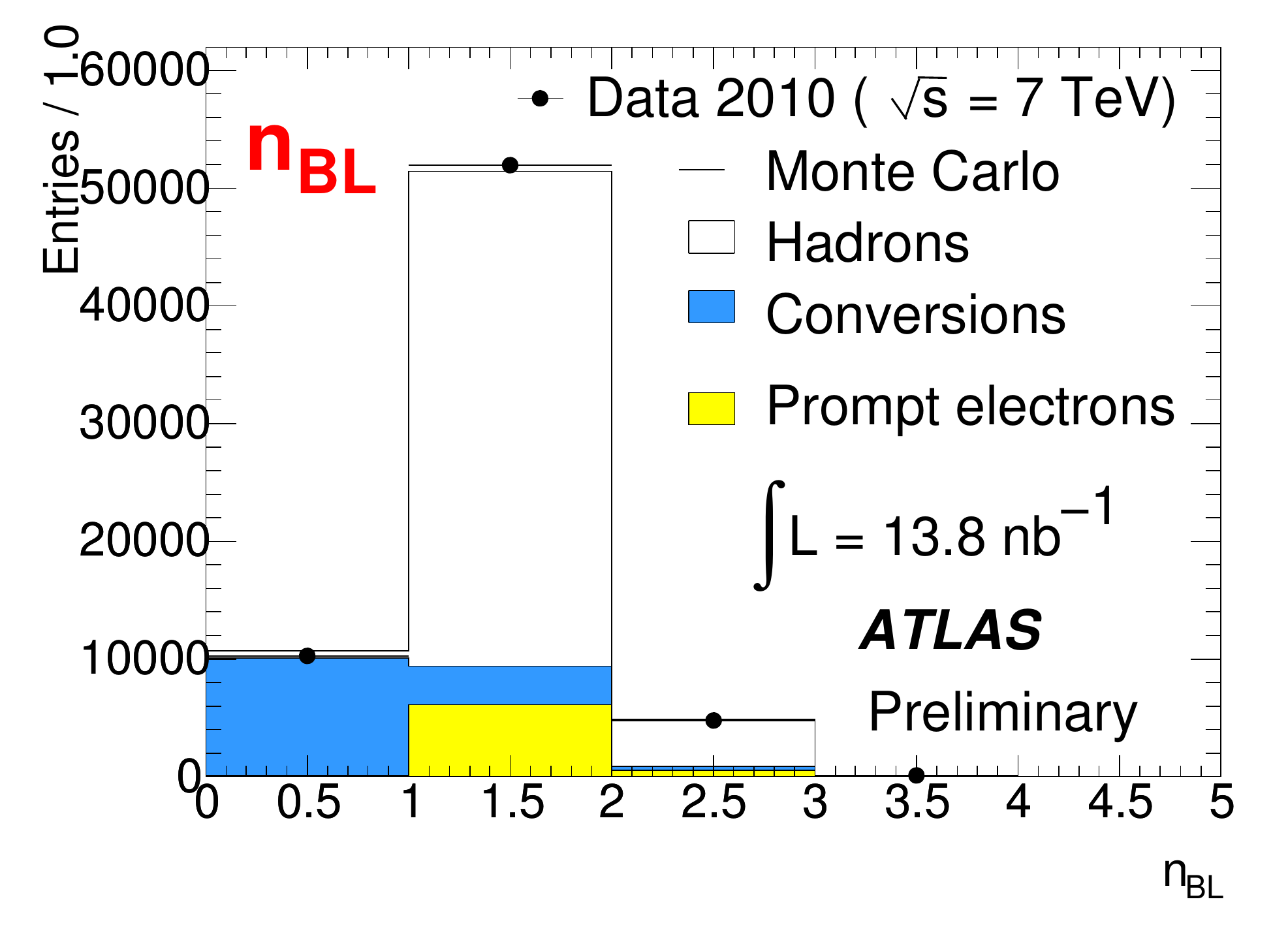}} 
  {\includegraphics[width=\figsize,height=3.2cm]{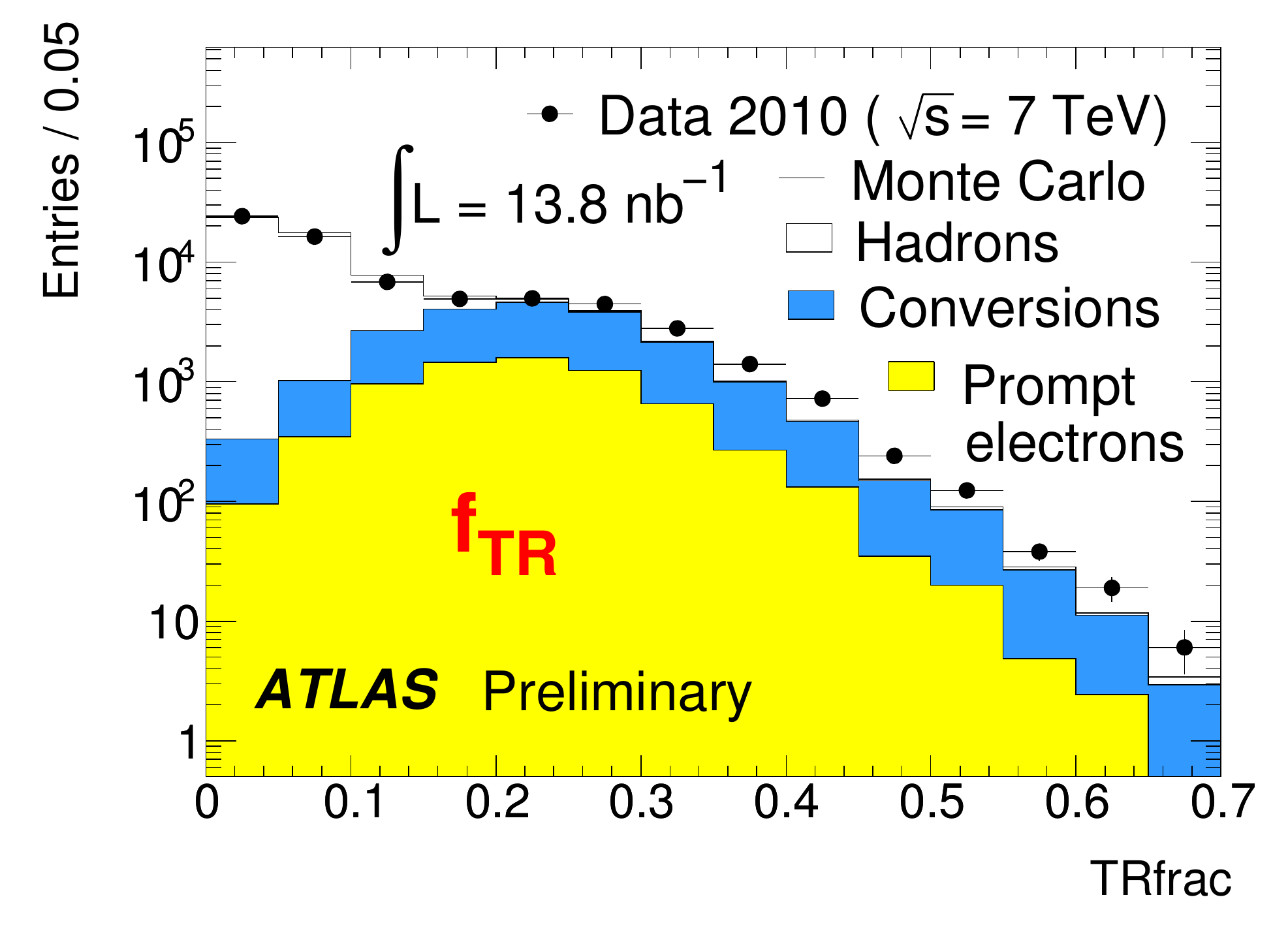}} 
  {\includegraphics[width=\figsize,height=3.2cm]{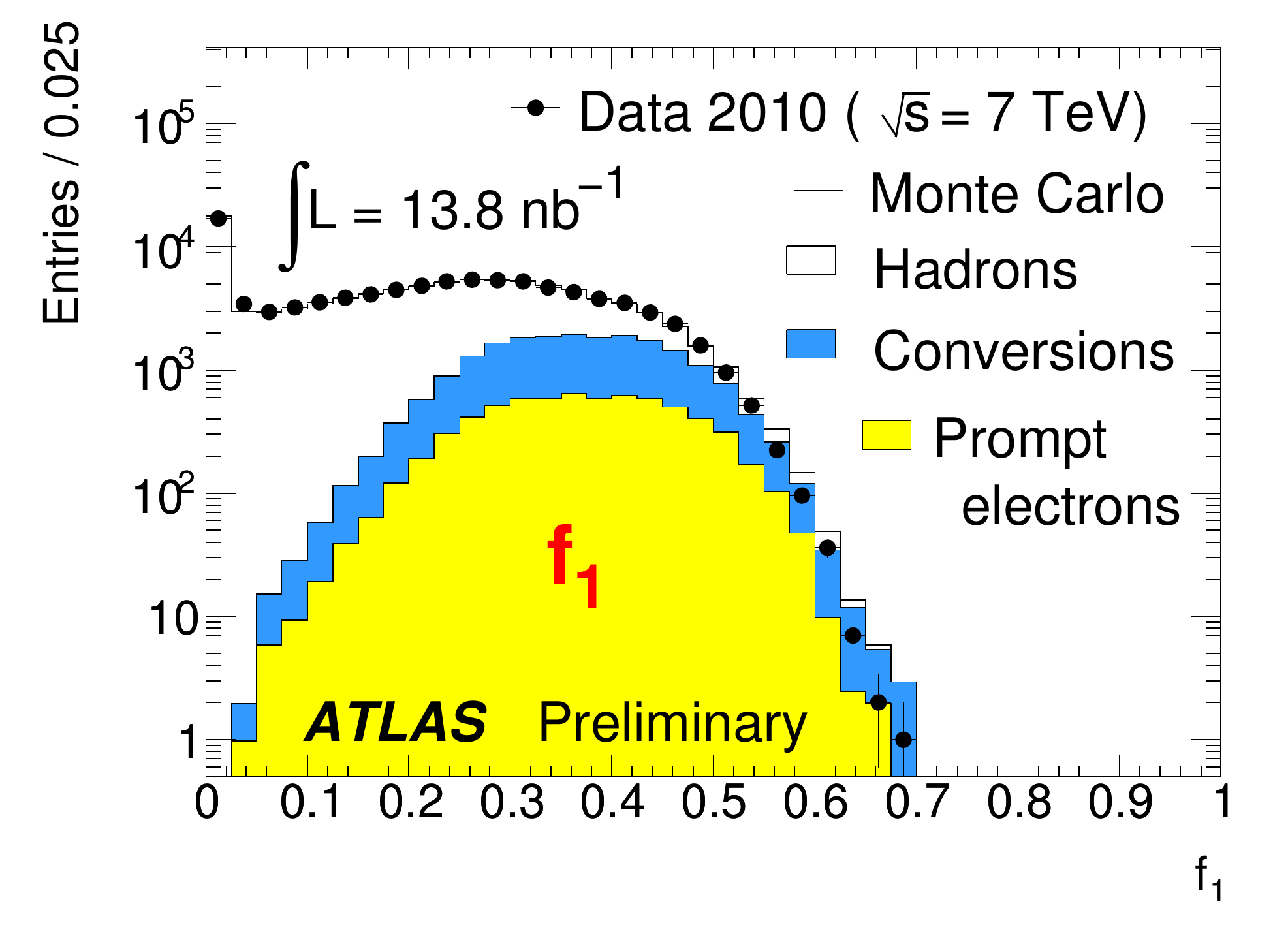}} 
  \vskip -0.3cm
  \caption{Inclusive electron analysis discriminant variables
    for data and non-diffractive minimum bias MC.}
  \vskip -0.5cm
  \label{fig:discrim}
\end{figure}

We decompose the sample using the ``matrix method,''
exploiting the differing efficiencies of each source
to pass selections on discriminating variables (\figref{fig:discrim}).
$\fTR$, the fraction of TRT
hits passing a high threshold, discriminates between
electrons and hadrons; $f_1$ is used
as a cross-check.  To discriminate
between electrons and conversions, we use $\nBL$, the number of hits
in the innermost (B) pixel layer; conversions usually occur
outside this layer.  
We find the
efficiencies for samples $Q$ and $\gamma$ using
Monte Carlo (MC) simulations; for $h$,
we use data with an inverted $f_1$ selection.
Our observed prompt electron signal (\tabref{tab:decomposition}, \figref{fig:prompt-electron}) is 
$9920\statsyst{160}{990}$ events.

\begin{table}
\twoboxes
{\small\begin{tabular}{|c|cc|} 
    \hline
                 & Data          & MC\\
    \hline
       $h$      & $43470\pm240$ & $46730\pm150$\\
       $\gamma$ & $13160\pm150$ & $13580\pm80$\\
       $Q$      & $9920\pm160$  & $6890\pm60$\\
       \hline
       Total    & 67124&\\
       \hline
     \end{tabular}%
  \caption{Decomposition of the inclusive electron sample by component,
  compared to \progname{pythia} \cite{pythia}.}%
  \label{tab:decomposition}}%
{\includegraphics[width=\figsize,height=3.4cm]{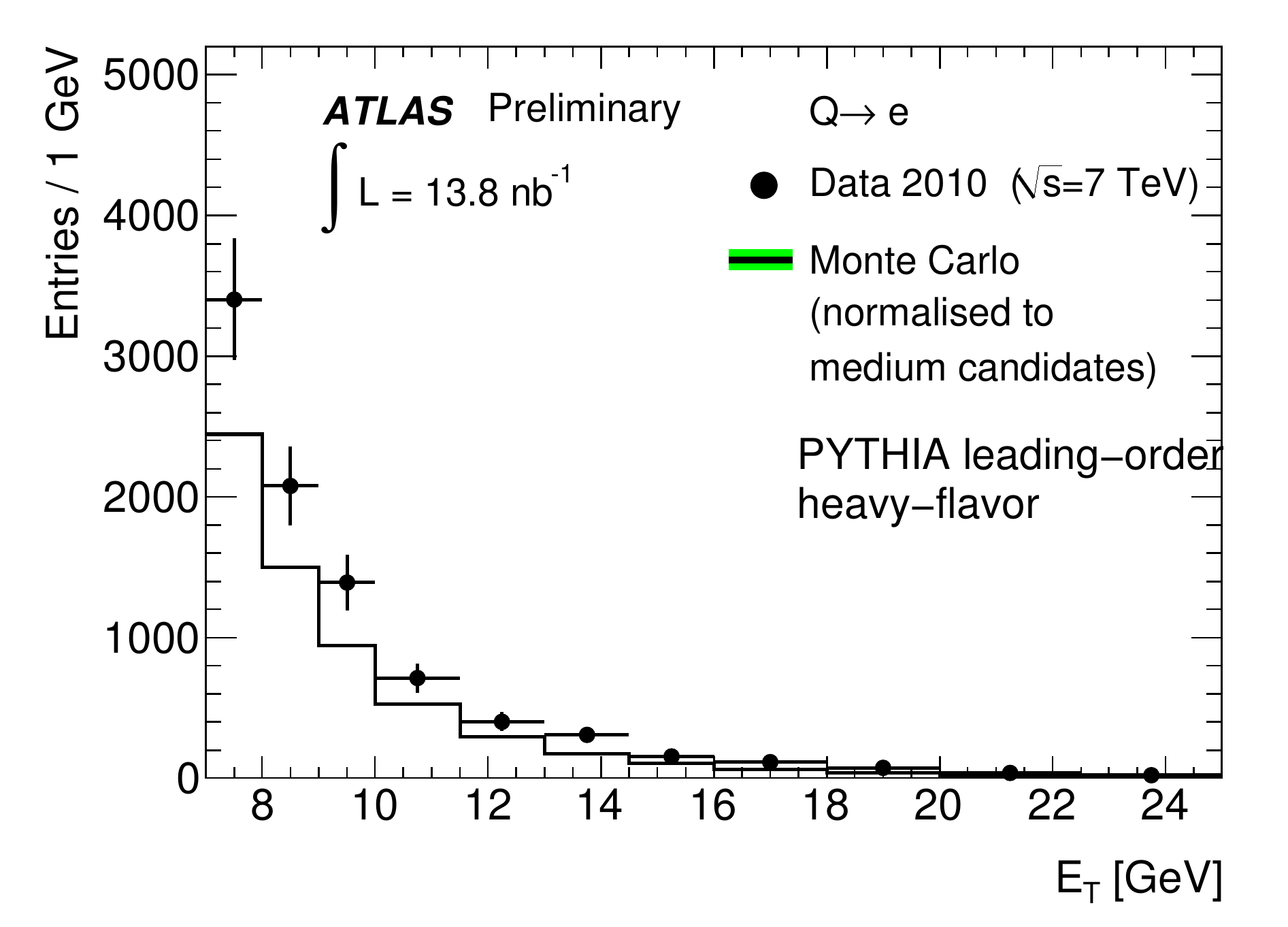} 
  \switchtofigure
  \vskip -0.5cm
  \caption{Prompt electron $\et$ spectrum.}%
  \label{fig:prompt-electron}}%
\end{table}



\section{Prompt photon analysis}

This analysis~\cite{promptphoton} reconstructs
prompt, isolated photons
and measures the purity of the sample.  Our initial selection
requires photon candidates with $\et>10\gev$ and $|\eta|<2.37$,
yielding 268992 events in $15.8\inb$.  We then
require small leakage out the back of the EM calorimeter,
small width in calorimeter layer~2, and large
$R_\eta = E(3\times7)/E(7\times7)$,
which also indicates a narrow cluster.

\begin{figure}
  \twoboxes
  {\includegraphics[width=\figsize,height=3.1cm]{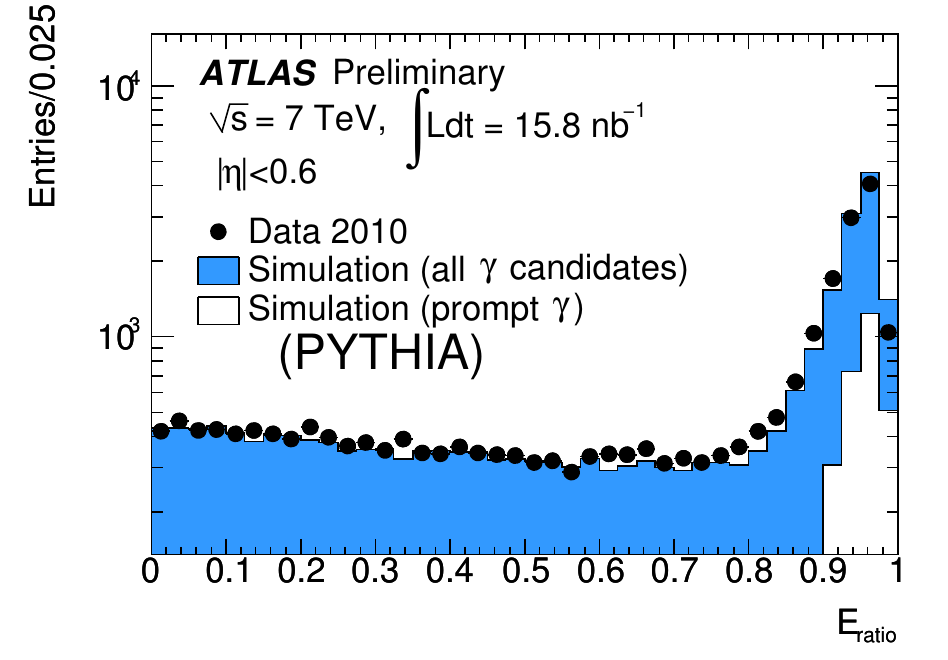}}%
  {\includegraphics[width=\figsize,height=3.1cm]{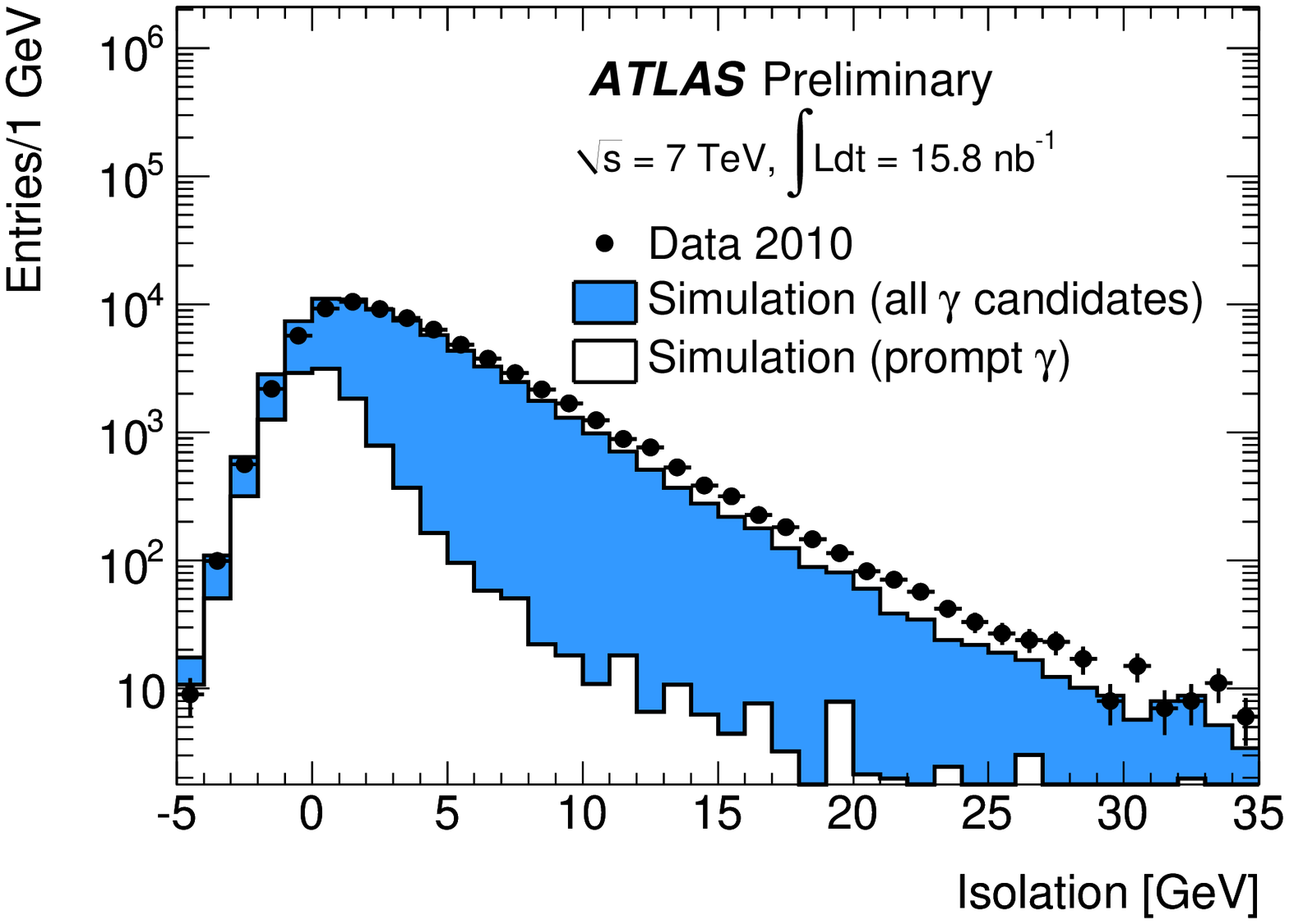}} 
  \label{fig:gamvars}%
  \vskip -0.3cm
  \caption{(left) $E_{\textrm{\scriptsize ratio}}$, the asymmetry between
  the first two maxima in layer~1.  (right) Photon isolation.}
\end{figure}

Two more requirements define our final sample.
First, tight cuts on the cluster shape and width
in layer~1 which select prompt photons and
reject $\pi^0$ decays.  
\Figref{fig:gamvars}, left, shows one such variable.
Second, isolation: the energy sum within
$R < 0.4$ around the candidate
must be $<3\gev$ (\figref{fig:gamvars}, right).
We find the sample purity by counting the number of candidates
that pass each of the four combinations of the cluster shape
and isolation selections.  Assuming that these selections
are uncorrelated and that the amount of prompt photon signal outside
of the signal region is negligible, the signal and background fractions
in the signal region may be calculated.  (We use simulations to
correct for the failure
of these assumptions to hold exactly.)
Results are in \tabref{tab:photonresults} and \figref{fig:photons}.
A prompt photon signal is seen for $\et > 15\gev$,
with purity over $70\%$ for $\et>20\gev$.

\begin{table}
  \twoboxes
  {\small\begin{tabular}{|l|l|l|l|} 
    \hline
    $\et$   & $N_{\textrm{\scriptsize cand}}$ & Purity (\%) & $N_{\textrm{\scriptsize sig}}$\\
    (GeV)&&&\\
    \hline
    10--15 & 5271 & $24\pm{5}\pm{24}$ & $1289\pm{297}\pm{1362}$\\
    15--20 & 1213 & $58\pm{5}\pm{8}$  &  $706\pm{69}\pm{86}$\\
    $>20$  &  864 & $72\pm{3}\pm{6}$  &  $618\pm{42}\pm{59}$\\
    \hline
  \end{tabular}%
  \caption{Total photon candidates and purity
  and number of signal events in the signal region.}%
  \label{tab:photonresults}}%
{\includegraphics[width=\figsize,height=3.4cm]{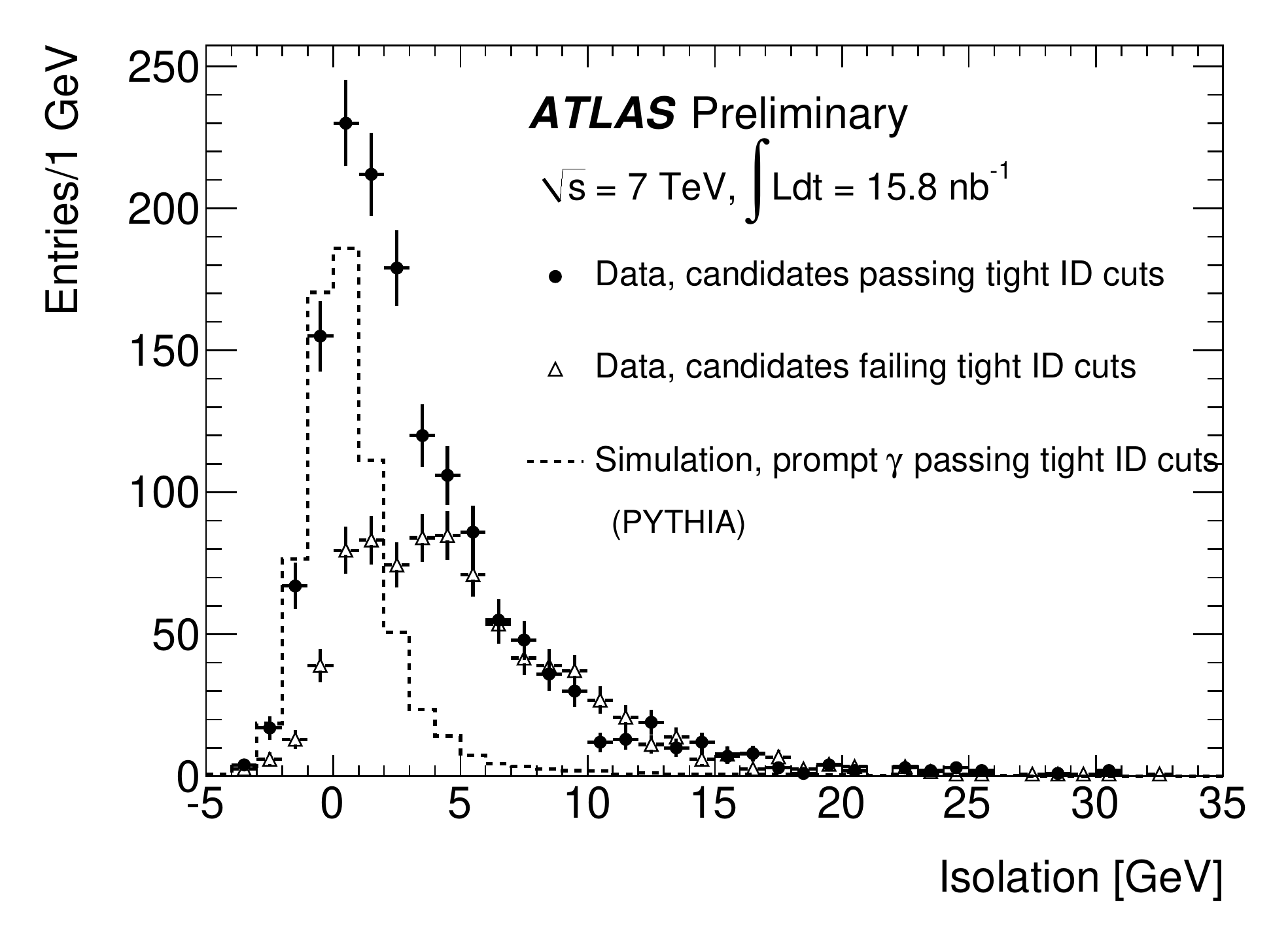} 
  \switchtofigure
  \vskip -0.5cm
  \caption{$\gamma$ candidate isolation, $\et>20\gev$.}%
  \label{fig:photons}}%
  \vskip -0.3cm
\end{table}

\section{J/$\psi$ analysis}

This analysis~\cite{jpsi} uses $77.8\inb$ and reconstructs a $J/\psi$ 
mass peak, using it to measure the shapes of some of
the discriminating variables.
For improved efficiency at low $\et$, this analysis uses
nearest-neighbor, rather than rectangular, cluster seeds.
We select opposite-sign pairs of electron candidates, one with
$\et>4\gev$ and one with $\et>2\gev$.  We make further selections
on $R_\eta$, $f_1$, the cluster's shape in layer~1,
and the track impact parameter, number of hits, and $\fTR$.

\begin{figure}
  \vskip -0.4cm
  \threeboxes
  {\includegraphics[width=\figsize,height=2.9cm]{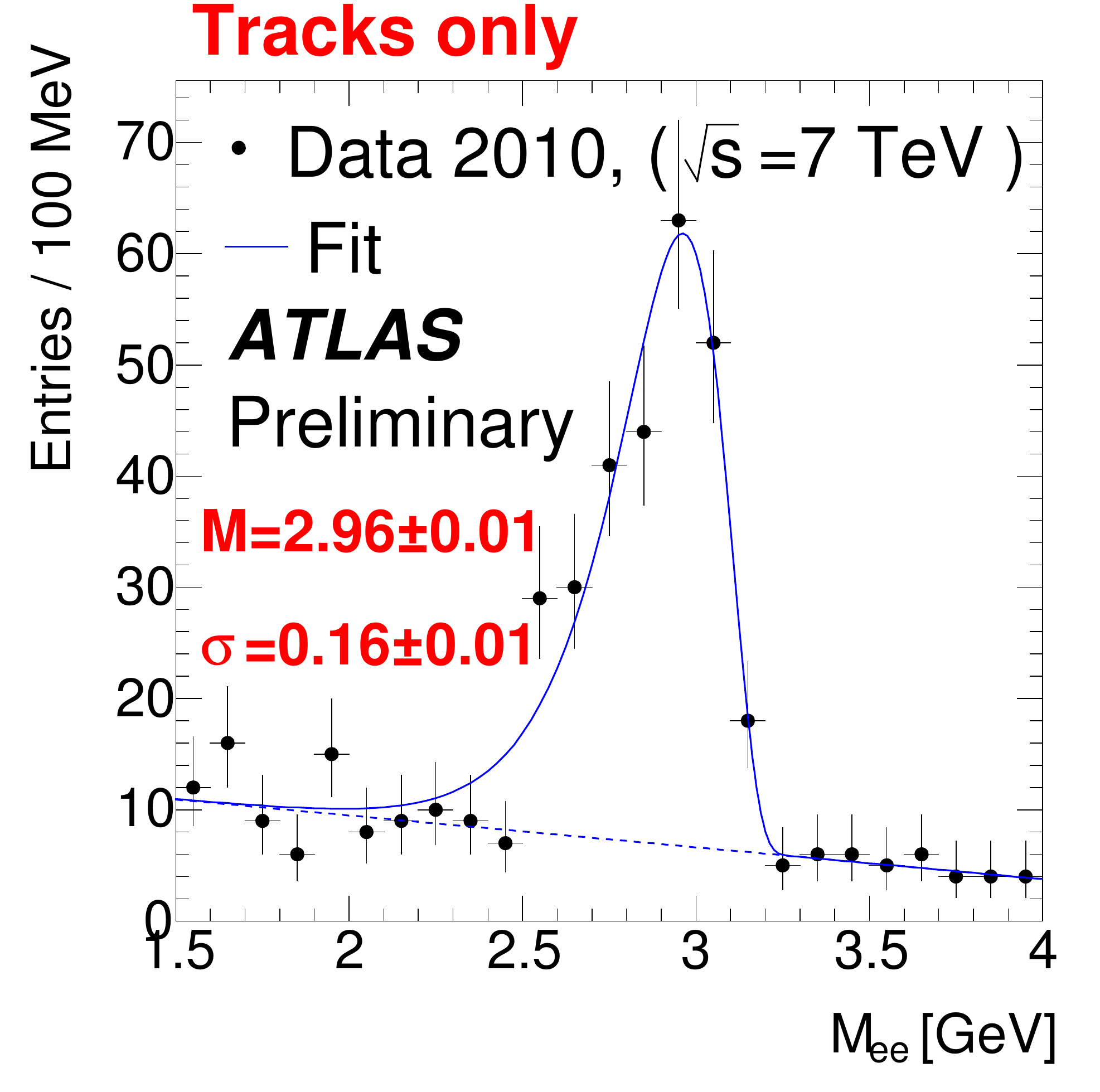}}
  {\includegraphics[width=\figsize,height=2.9cm]{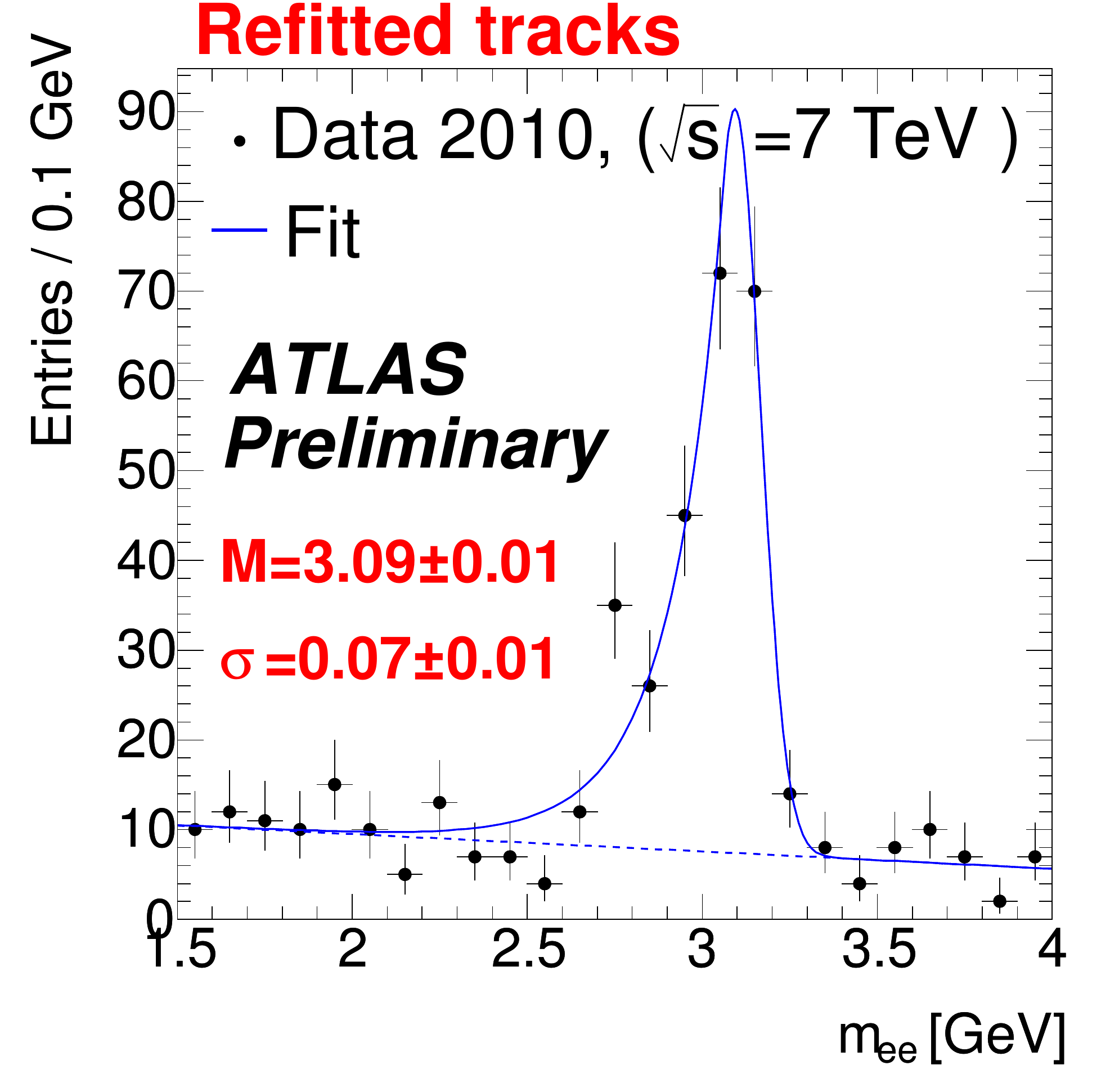}} 
  {\includegraphics[width=\figsize,height=2.9cm]{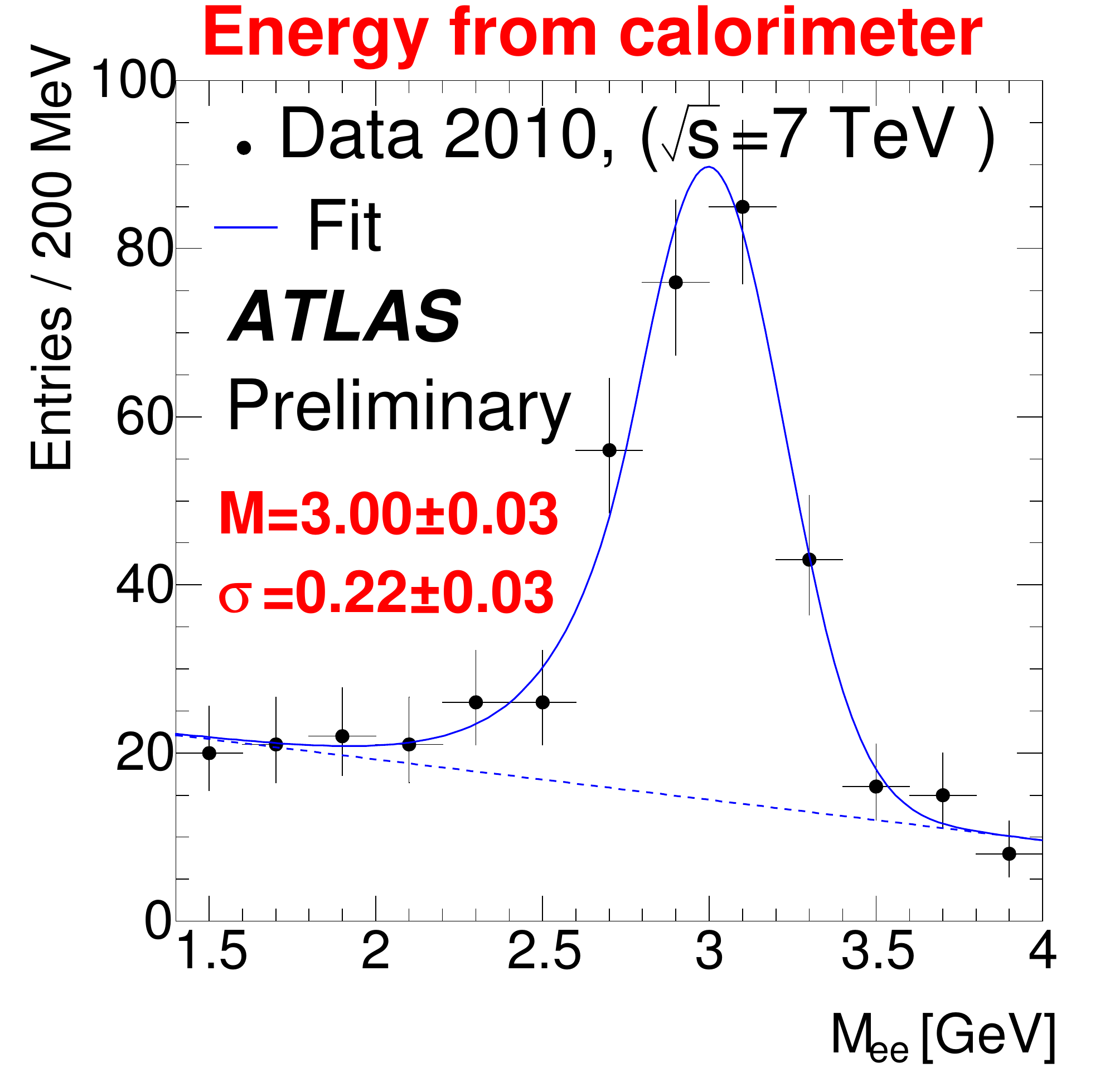}} 
  \vskip -0.3cm
  \caption{$J/\psi$ mass peak reconstructed with three different methods.}
  \label{fig:mjpsi}
\end{figure}

We calculate the $J/\psi$ mass in three ways.
First, using kinematics from
tracks only (\figref{fig:mjpsi}, left).
This gives a mass of $2.96\pm0.01\gev$, slightly low compared
to the Particle Data Group value of $3.097\gev$~\cite{pdg}.  This is expected,
since this method ignores energy losses due to bremsstrahlung.
Second, taking into account the energy loss by bremsstrahlung
in the tracker material.
This (\figref{fig:mjpsi}, middle) yields $3.09\pm0.01\gev$.
Third, taking energies from the calorimeter clusters
and directions from the tracks.  This gives (\figref{fig:mjpsi}, right)
$3.00\pm0.03\gev$, also slightly low, as the current
calorimeter energy calibrations are known to be suboptimal
at these low cluster energies.

\begin{figure}
  \vskip -0.3cm
  \threeboxes
  {\includegraphics[width=\figsize,height=2.9cm]{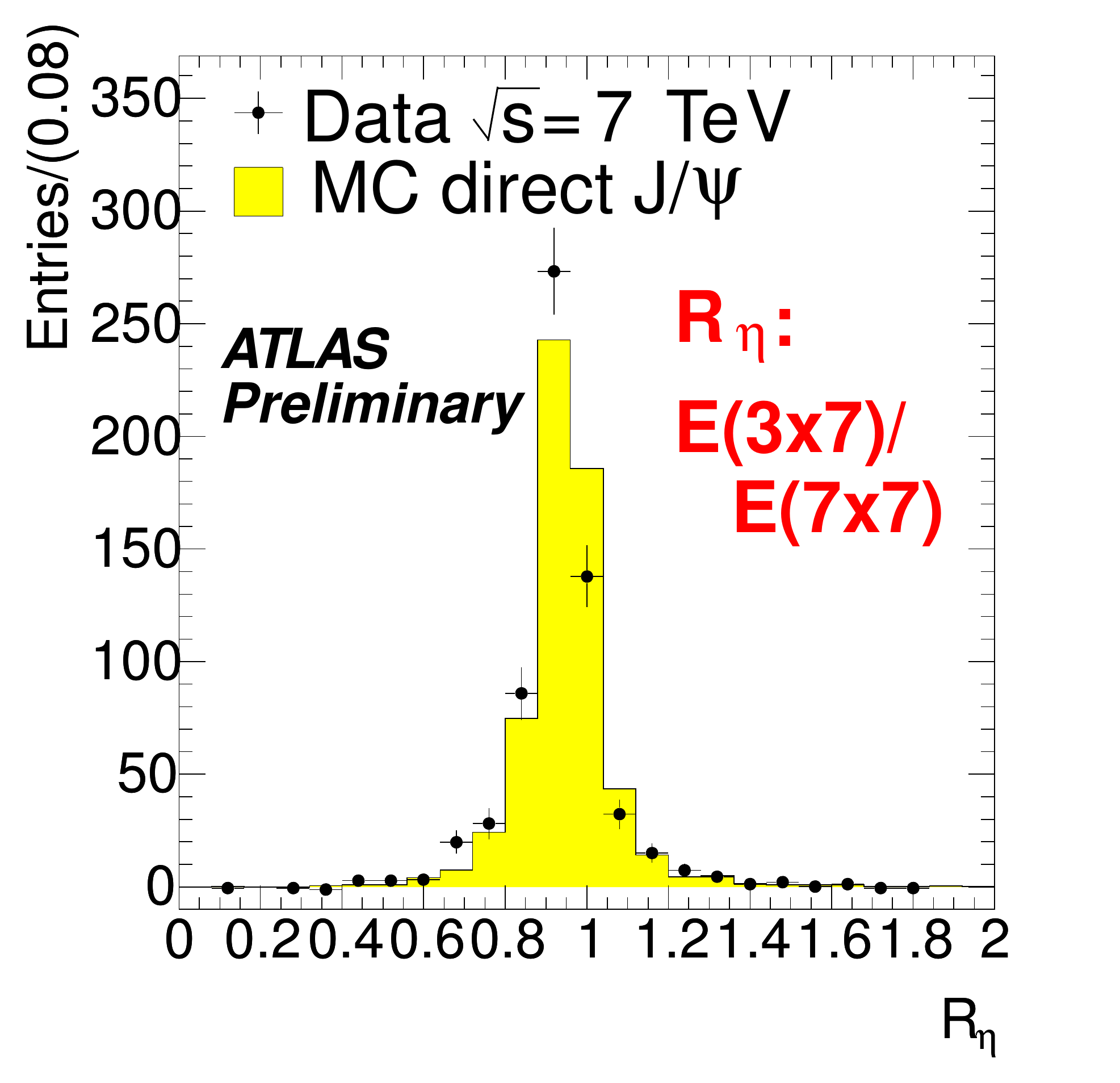}}
  {\includegraphics[width=\figsize,height=2.9cm]{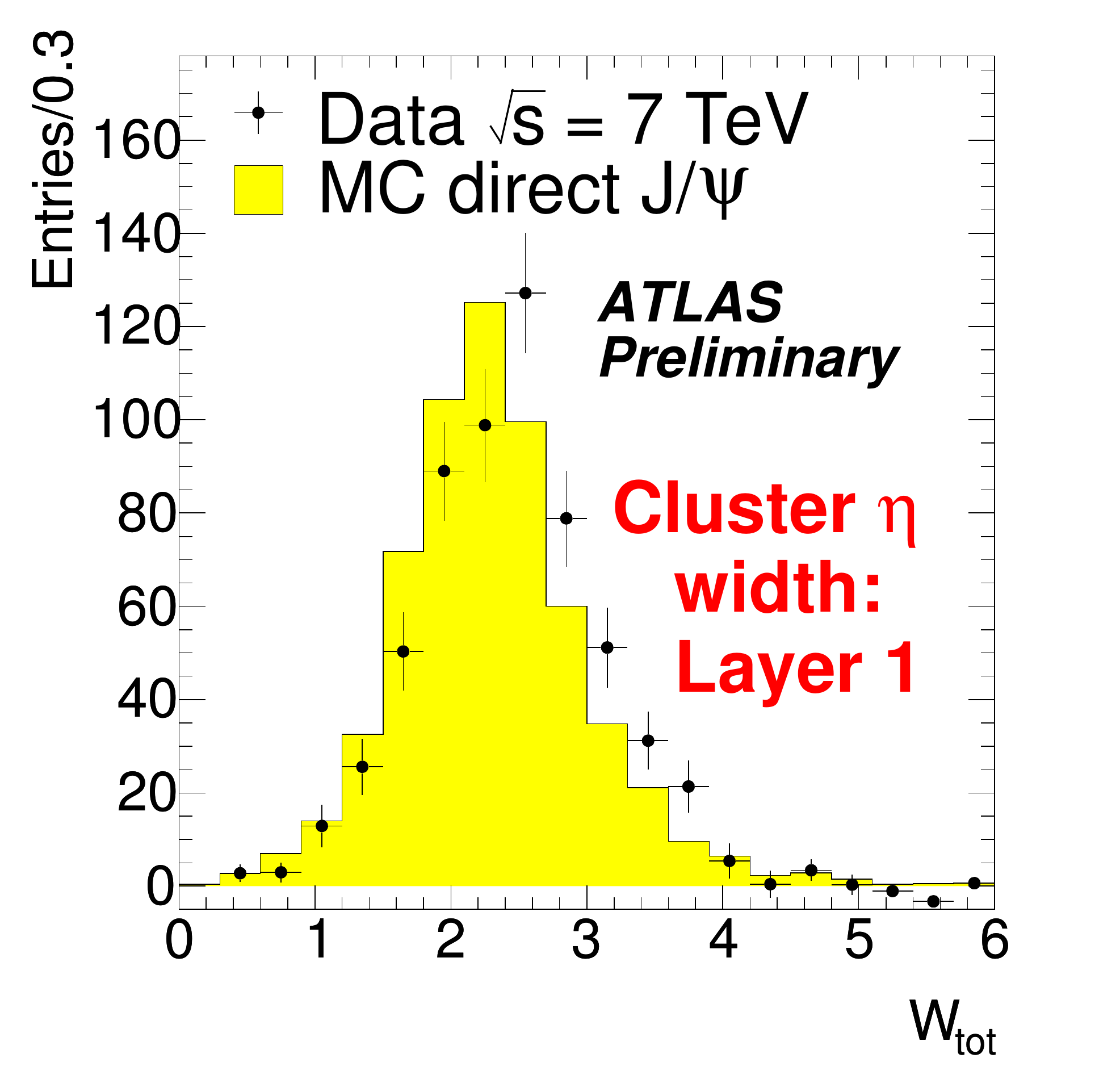}}
  {\includegraphics[width=\figsize,height=2.9cm]{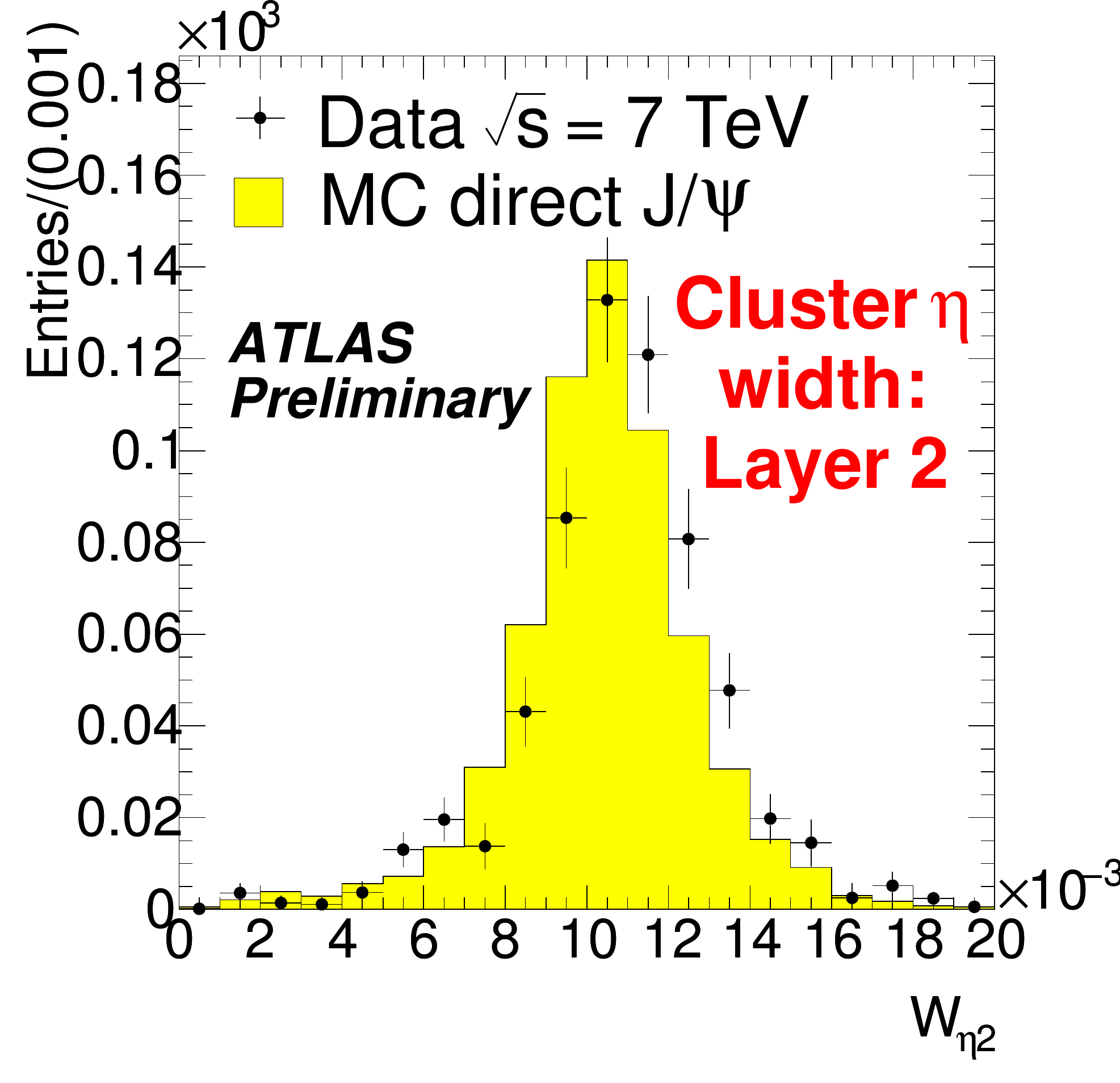}}
  \vskip -0.3cm
  \caption{Electron discriminant variables from electrons from $J/\psi$ decay,
  compared to \progname{pythia}.}
  \vskip -0.3cm
  \label{fig:jpsi-tp}
\end{figure}

The $J/\psi$ peak
defines real electrons that can be used to check
the detector simulation.
We perform a ``tag-and-probe'' analysis: we maintain
tight selections on one, \emph{tag}, candidate, and remove
the shower shape selections from the other, \emph{probe}, candidate.
We then select candidate pairs with masses within 2.7--$3.2\gev$
and $f_1>0.15$ and plot variables from the probe.
There is general agreement between data and the
simulation (\figref{fig:jpsi-tp}); however, statistics
now available reveal some small systematic differences in lateral shower shapes.
Work is in progress on understanding these.

\section{Conclusions}

ATLAS and the LHC are performing well;
the luminosity is increasing rapidly, and we will soon have
large electron samples from both $J/\psi$
and $W/Z$~decays.  We are working
to better understand the detector performance
in preparation for new discoveries
in electron/photon channels.

This work is supported in part by the U.S.~Department of Energy
under contract
DE-AC02-98CH10886 with Brookhaven National Laboratory.

\end{document}